\documentclass[aps,pre,twocolumn,groupedaddress,amsmath,amssymb,showpacs]{revtex4}
\usepackage{graphicx}

\usepackage{subfigure}

\begin{document}

%Title of paper
\title{Statistical mechanics of inference}
\author{Jonathan Landy}
\email{landy@mrl.ucsb.edu}
\affiliation{Materials Department, University of California, Santa Barbara}

\date{\today}

\begin{abstract}
Statistical modeling often involves identifying an optimal estimate to some underlying probability distribution  known to satisfy some given constraints.  I show here that choosing as estimate the centroid, or center of mass, of the set consistent with the  constraints formally minimizes an objective measure of the expected error.  Further, I obtain  a useful approximation to this point, valid in the thermodynamic limit, that immediately provides much information relating to the full solution set's geometry.  For weak constraints, the centroid is close to the popular maximum entropy solution, whereas for strong constraints the two are far apart. Because of this, centroid inference is often substantially more accurate. The results I present allow for its straightforward application.
\end{abstract}

\maketitle

One is sometimes confronted with the challenge of estimating probabilities from partial information.  For example, given a stochastic system that transitions between a very large number of distinct states, the sampling time required to directly obtain a statistically significant estimate -- through binning, say -- to the occupation probability of some particular state may be prohibitively long.  This is often the case in neuroscience experiments, because the number of distinct states that a neural network can access grows exponentially with network size \cite{ Shl-06, Sch-06, coc-09}.  
Although rigorous distribution identification is not possible in such situations, inference strategies that intelligently make use of available data can provide good estimates \cite{Pre-13}.
Here, I consider probabilistic inference in the uniform ensemble, where all distributions consistent with a given set of constraints are supposed equally likely.  Using methods of statistical mechanics, I obtain a simple approximation to the centroid of the solution set, defined  by equations (\ref{saddle_conds})-(\ref{centroid_est}), below.  This, in turn, leads to useful results characterizing the full solution set's geometry, and it also allows for comparison to the maximum entropy solution. I find that the centroid is sometimes expected to be substantially more accurate.

I consider here the following general scenario:  It is given that a desired, underlying distribution $\textbf{p}^* \equiv (p_1, p_2, \ldots, p_N)$ on $N$ states, with $p_i^* \in [0,1]$ the probability of state $i$,
satisfies a set of $\mathcal{C} \ll N $ linear constraints of the form
\begin{eqnarray}\label{constraints}
\sum_i p_i^* f_{ji} = 1, \ \ \ j \in \{1,2,\ldots, \mathcal{C} \},
\end{eqnarray}
with the real-valued $\left \{f_{ji} \right \}$ specified.  The distribution is normalized to
\begin{eqnarray}\label{normalization}
\sum_i p_i^*= 1.
\end{eqnarray}
 I refer to the set $\mathcal{S}$ of distributions $\textbf{p}$ satisfying (\ref{constraints}) and (\ref{normalization}) as the solution set.  We are  to select from $\mathcal{S}$ one distribution that is optimal:  Here, I consider the case where the selected distribution is supposed to be a good approximation to the unknown $\textbf{p}^*$.  I take as a measure of error in estimate $\textbf{p}$ the quantity
 \begin{eqnarray} \label{error}
E \left (\textbf{p}^*, \textbf{p} \right) \equiv \left \vert\textbf{p}^* - \textbf{p} \right \vert ^2 = \sum_i \left  (p_{i}^* - p_{i} \right)^2,
\end{eqnarray}
 the squared distance between the underlying and the estimated distributions.
I stress that (\ref{error}) is not necessarily the only appropriate measure of error in an inference problem.  However, it does represent an objective measure that is both familiar and useful to consider.
 
In certain situations, it may be appropriate to consider certain members of $\mathcal{S}$ more likely to be $\textbf{p}^*$ than others.   For example, if we know that $p^*$ was generated by a process more likely to generate sparse distributions, sparse members of $\mathcal{S}$ should be weighted more heavily \cite{alb-12}.  However, in the absence of such information,  an axiom of equal probability is appropriate:  Every state consistent with (\ref{constraints}) and (\ref{normalization}) should be considered equally likely to be the underlying distribution \footnote{This is an application of Laplace's rule of indifference.}.  I work under this axiom here.  In this case, the expected error in an estimate $\textbf{p}$ is obtained by averaging  (\ref{error}) over $\textbf{p}^*$,
\begin{eqnarray} \label{av_error}
 \left \langle E(\textbf{p}^*,\textbf{p}) \right \rangle_{\textbf{p}^*\in \mathcal{S}} = \sum_i p_i^2 - 2 p_i \left \langle p_i^* \right \rangle_{\mathcal{S}} + \left \langle p_i^{*2} \right \rangle_{\mathcal{S}}.
\end{eqnarray}
The solution $\textbf{p}^c$ that minimizes the expected error  is obtained by setting the derivative of (\ref{av_error}), with respect to $p_i$, to zero.   This gives,
\begin{eqnarray}
\textbf{p} \to \textbf{p}^c  = \left \langle \textbf{p}^* \right \rangle_{\mathcal{S}},
\end{eqnarray}
the centroid of the solution set \footnote{It is a simple matter to prove that $\mathcal{S}$ is convex.  Thus, $\textbf{p}^c \in \mathcal{S}$:  The centroid is  a valid solution.}.  This represents the formal solution to a particular, well-defined inference problem.  Namely, this returns the distribution in $\mathcal{S}$ minimizing (\ref{av_error}).   Unfortunately, a simple, general formula for $\textbf{p}^c$ does not exist \cite{rad-07}.    However, in the following, I obtain an estimate to $\textbf{p}^c$ that is easy to evaluate.  Comparison to this $\textbf{p}^c$ estimate  then provides a simple method for testing the expected performance of other solutions:  For $\textbf{p}$ close to $\textbf{p}^c$, the expected error (\ref{av_error}) is nearly minimized.  On the other hand, the expected error (\ref{av_error}) is relatively large for $\textbf{p}$ far from $\textbf{p}^c$.

In order to characterize the solution set, I consider the configuration partition sum associated with a free particle, with position $\textbf{p}$, moving through $\mathcal{S}$. This is
\begin{eqnarray} \label{partition}
\mathcal{Z} = \int_0^{\infty}   \delta \left ( \sum p_i - s \right ) \prod_{j=1}^{\mathcal{C}}  \delta \left (  \sum_i p_i f_{ji} - t_j\right ) \prod_{i=1}^N dp_i,
\end{eqnarray}
where I have generalized slightly the constraint equations (\ref{constraints}) and (\ref{normalization}), now requiring the sum over probabilities to be equal to $s$ and the dot product of $\textbf{p}$ along $\textbf{f}_j$ to be $t_j$.   As defined, $\mathcal{Z}$ is simply equal to the volume of the solution set $\mathcal{S}$. The Laplace transform of $\mathcal{Z}$ is
\begin{eqnarray} \nonumber \label{partition2}
\tilde{Z} \left (m,\{\lambda_j \} \right) &\equiv& \int_{0}^{\infty} \mathcal{Z} \left (s, \{t_j \} \right) e^{- s m - \sum_j t_j \lambda_j} ds \prod_j d t_j \\
&=& \int_0^{\infty} e^{- m \sum p_i - \sum_i p_i \sum_j \lambda_j f_{ji}}\prod_{i=1}^N dp_i.
\end{eqnarray}
Here, I have used (\ref{partition}) to obtain the second line.  The  integrals over the $\{p_i\}$ are now decoupled, and $\tilde{Z}$ can be evaluated in closed form as
\begin{eqnarray}
\tilde{Z} \left (m,\{\lambda_j \} \right)  = \prod_{i=1}^N \frac {1}{m + \sum_j \lambda_j f_{ji}}.
\end{eqnarray}
The solution set volume is given formally by the inverse Laplace transform of this quantity,
\begin{eqnarray} \label{partition3}
\mathcal{Z} \left (s, \{t_j \} \right) \equiv \oint \tilde{Z} \left (m,\{\lambda_j \} \right)  e^{s m + \sum_j t_j \lambda_j} dm \prod_j d \lambda_j,
\end{eqnarray}
where the indicated contours are parallel to the imaginary axis \cite{Jon-43}.

In order to evaluate the integral (\ref{partition3}), I now assume that $N$, the number of accessible states, or components of $\textbf{p}$, is large.  In this case, the integrand in (\ref{partition3}) will be highly peaked, and an asymptotic series for $\log \mathcal{Z}$ can be obtained, the first term being the saddle point value \cite{Jon-43}.  Setting $s$ and the $\{t_j \}$ to their common, physical value, one, we have
\begin{eqnarray}\nonumber \label{saddle1}
 \mathcal{Z}   &=& \oint_{m, \{ \lambda_j \}} e^{m + \sum_j \lambda_j  
    - \sum_i \log \left ( m + \sum_j \lambda_j f_{ji}\right)} \\
 & \approx &  e^{m^* + \sum_j \lambda_j^*  
    - \sum_i \log \left ( m^* + \sum_j \lambda_j^* f_{ji}\right)},
\end{eqnarray}
where the saddle point $m^*$ and $\{\lambda_j^*\}$ values are those that leave the derivative of $\log \mathcal{Z}$ stationary. That is, they satisfy the following equations, obtained by setting the derivatives of the exponent in (\ref{saddle1}), with respect to $m$ and the $\{\lambda_j\}$, individually to zero:
\begin{eqnarray}  \label{saddle_conds}
1 - \sum_{i=1}^N \frac{1}{m^* + \sum_j \lambda_j^* f_{ji}} &=& 0 \\ \nonumber
1 - \sum_{i =1}^N \frac{f_{ji}}{m^* + \sum_j \lambda_j^{*} f_{ji}} &=& 0, \ \ \ j \in \{1, 2, \ldots \mathcal{C} \}.
\end{eqnarray}
We can solve directly for one unknown:  Summing over the second line above, multiplied by $\lambda_j^*$, and adding to this $m^*$ times the first line, gives
\begin{eqnarray}\label{m_sol}
m^* = N - \sum_j \lambda_j^*,
\end{eqnarray}
a  simple relationship.
The remaining $\mathcal{C}$ unknowns, the $\left \{ \lambda_j^* \right \}$, must be  solved for using (\ref{constraints}), or, equivalently, the latter  conditions of (\ref{saddle_conds}).

Notice that if we define
\begin{eqnarray} \label{centroid_est}
p_i^{c,1} \equiv \frac{1}{m^* + \sum_j \lambda_j^* f_{ji}},
\end{eqnarray}
the saddle point conditions (\ref{saddle_conds}) imply that the distribution $\textbf{p}^{c,1}$ satisfies both (\ref{constraints}) and (\ref{normalization}).    In fact, $\textbf{p}^{c,1}$ is the first-order, saddle point estimate to the centroid of $\mathcal{S}$.  This is most easily proven by introducing a field $h_i$ in (\ref{partition}) coupled to $p_i$.  Following steps similar to those shown above, this gives
\begin{eqnarray} \nonumber \label{partition4}
\mathcal{Z} &\to & \int   \delta \left ( \sum p_i - s \right ) \prod_{j}  \delta \left (  \sum_i p_i f_{ji} - t_j\right ) e^{-\sum_i h_i p_i}  \\
&=& \oint_{m, \{ \lambda_j \}} e^{m + \sum_j \lambda_j  
    - \sum_i \log \left ( m + \sum_j \lambda_j f_{ji} + h_i\right)}.
\end{eqnarray}
From the first line above, we obtain
\begin{eqnarray} \label{gen_cent}
p^c_i \equiv 
\langle p_i \rangle_{\mathcal{S}} =-\left . \partial_{h_i}  \log  \mathcal{Z} \right \vert_{\{h_j\} = 0},
\end{eqnarray}
an exact identity.
Applying the saddle point approximation to (\ref{partition4}) gives the analog of (\ref{saddle1}), with the $h_i$ field included.  Plugging in to (\ref{gen_cent}) then gives $\textbf{p}^c \sim \textbf{p}^{c,1}$, the value in (\ref{centroid_est}).  More accurate estimates are obtained through expansion about the saddle point.  For example, writing $m = m^* + \delta m$ and $\lambda_j = \lambda_j^* + \delta \lambda_j$, evaluation of the Gaussian fluctuations about the saddle point gives
\begin{eqnarray} \nonumber \label{centroid_est_2}
\log \mathcal{Z} &\sim& \sum_{j=0}^{\mathcal{C}} \lambda_j^* - \sum_i \log \left ( h_i+ \sum_{j=0}^{\mathcal{C}} \lambda_j^* f_{ji}  \right) \\
&&- \frac{1}{2} \log \det \mathcal{M},
\end{eqnarray}
where $\mathcal{M}$ is the $(\mathcal{C}+1) \times (\mathcal{C}+1)$ matrix with components
\begin{eqnarray}
\mathcal{M}_{\alpha \beta} \equiv \sum_i \frac{f_{\alpha i} f_{\beta i}}{\left (h_i + \sum_{j=0}^{\mathcal{C}} \lambda_j^* f_{ji} \right)^2}.
\end{eqnarray}
Here, I have written $m^* \equiv \lambda^*_0$ and $f_{0i} \equiv 1$, in order to briefly simplify notation.
Combining  (\ref{gen_cent}) and (\ref{centroid_est_2}) gives the second order centroid estimate $\textbf{p}^{c,2}$, a refinement to $\textbf{p}^{c,1}$.  In order to carry out the implied variation of (\ref{centroid_est_2}) with respect to $h_i$ here, the field dependences of the $\left \{ \lambda_j^* \right \}$ are  needed within $\mathcal{M}$ \footnote{In evaluating $\textbf{p}^{c,1}$ this is not necessary because the $ \left \{\lambda_j^* \right \}$ leave the exponent stationary at the saddle point level.}.  Differentiating the saddle point equations, $\sum_i f_{ji} \left (h_i+\sum_k \lambda_k^* f_{ki} \right)^{-1}  =1$, gives the matrix equation
\begin{eqnarray}
\mathcal{M}_{\alpha \beta} \partial_{h_i} \lambda_{\beta}^* =  - f_{\alpha i} \left( p^{c,1}_i \right)^2,
\end{eqnarray}
which can be inverted to solve for the necessary derivatives.
Carrying out this procedure is useful for small $N$.   However, for $N \gtrsim 100$, $\textbf{p}^{c,1}$ already provides an accurate  approximation to $\textbf{p}^c$.

Once $\textbf{p}^{c,1}$ has been evaluated, one can immediately characterize, approximately, the solution set's geometry.  For example, from (\ref{partition4}), the variance of $p_i$ is 
\begin{eqnarray}
\sigma^2_{p_i} \equiv \left \langle p_i^2 \right \rangle_{\mathcal{S}} - \left \langle p_i \right \rangle_{\mathcal{S}}^2  = \left . \partial^2_{h_i}  \log  \mathcal{Z} \right \vert_{\{h_j\} = 0}.
\end{eqnarray}
Plugging in the saddle point estimate for $\mathcal{Z}$ gives
\begin{eqnarray}\label{width}
\sigma_{p_i} \sim p_i^{c,1}.
\end{eqnarray}
That is, the width of the solution set in the $\hat{\textbf{e}}_i$ direction is approximately equal to $p_i^c$, the $i^{\text{th}}$ component of the solution set's centroid.  Higher-order cumulant averages also immediately follow. Further, at the saddle point level, from (\ref{saddle1}), (\ref{m_sol}), and (\ref{centroid_est}),
\begin{eqnarray}
\mathcal{Z} \sim e^N \prod_{i=1}^N \frac{1}{m^* + \sum_j \lambda_j^* f_{ji}} 
=  e^N \prod_{i=1}^N p^{c,1}_i.
\end{eqnarray}
We see that the solution set volume is proportional to the product of the centroid's components.  This provides a simple, qualitative means for determining whether a given set of constraints (\ref{constraints}) is strong or weak:  By the arithmetic-geometric mean inequality, we have
\begin{eqnarray} \nonumber \label{a-gmeaninq}
\mathcal{Z} \sim  e^N \prod_{i=1}^N p^{c,1}_i &\leq & e^{N} \times \left \{\frac{1}{N} \sum_i p^c_i\right\}^{N} \\
&=& \exp \left [N - N \log N \right],
\end{eqnarray}
where I have made use of the normalization condition (\ref{normalization}) to obtain the second line.
  Equality holds here if and only if each of the $\left \{p_i^{c,1}\right \}$ are equal to $\frac{1}{N}$, which is the case only in the absence of  constraints.  If constraints are applied, and the $\{p_i^{c,1} \}$ are substantially different in magnitude, the upper bound in (\ref{a-gmeaninq}) is far from strict.  In this limit, the solution set volume is significantly diminished, and the constraints can be considered strong.  On the other hand, if the $\{p_i^{c,1} \}$ are all similar in magnitude, $\log \mathcal{Z}$ is only slightly diminished, and the constraints can be considered weak.

\begin{figure}[h]
\begin{center}\scalebox{.12}
{\includegraphics{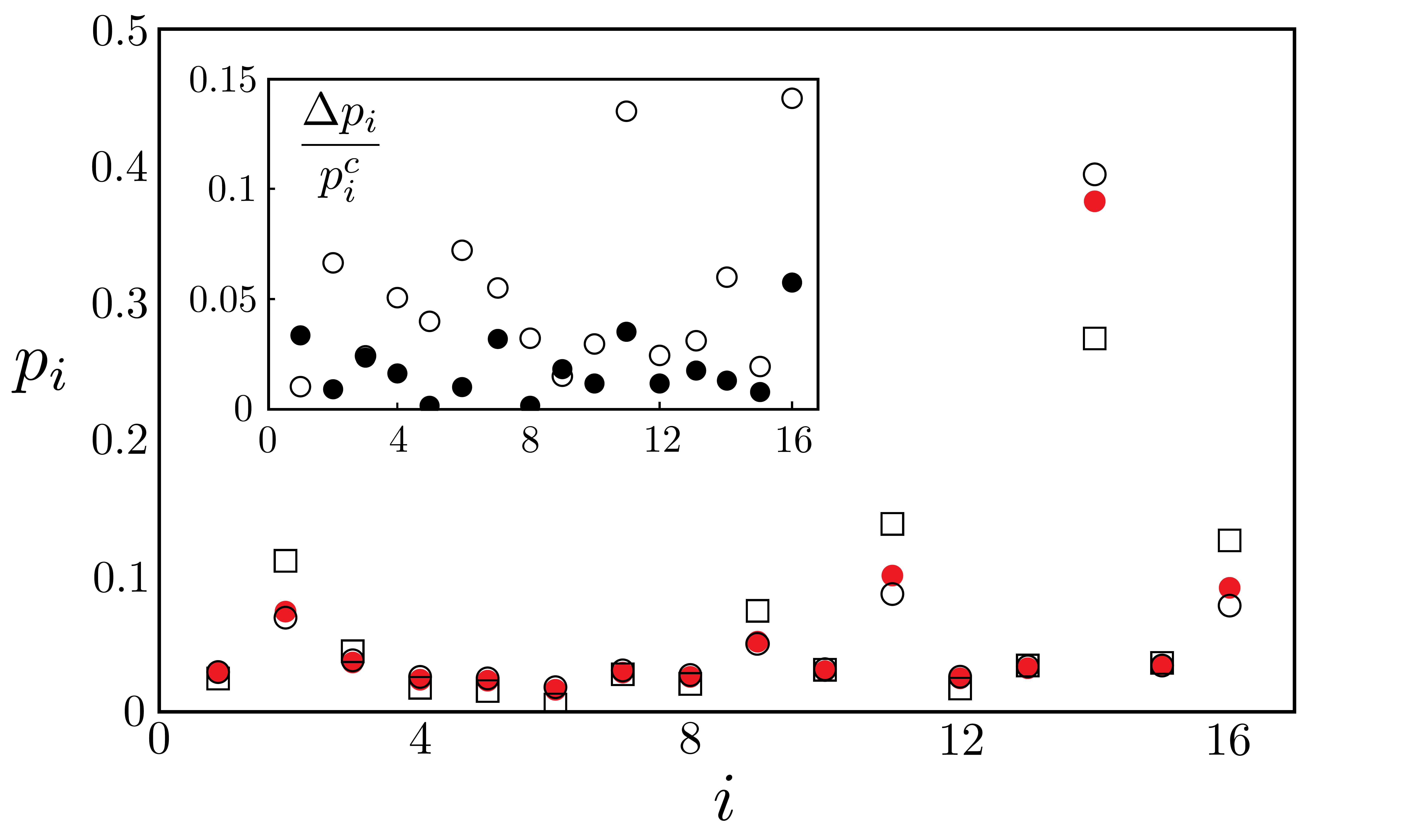}}
\caption{\label{fig:snub}
(color online) Centroid $p_i^{c}$ (red, filled circle),  centroid estimate $p_i^{c,1}$ (open circle), and maximum entropy solution $p_i^{ME}$ (square) versus index $i$ for an $N=16$ system, subject to a single, strong constraint $\textbf{p} \cdot \textbf{f} =1$.  Here, the components of $\textbf{f}$ were selected at random, with $P(f_i) \propto \exp[- f_i^2 /2]$.  The centroid $\textbf{p}^c$ was obtained by averaging over a $10^9$ step random walk through  $\mathcal{S}$.     (inset):  Relative absolute error in centroid estimates:  $\vert p_i^{c,j} - p_i^c \vert /p_i^c$, for $j =1$ (open) and $j=2$ (filled).  
}
\end{center}
\end{figure}

We are now in a position to compare the centroid solution $\textbf{p}^c$, which minimizes the expected error (\ref{av_error}), to the maximum entropy solution $\textbf{p}^{ME}$, which maximizes 
\begin{eqnarray}
S \equiv -\sum_i p_i \log p_i,
\end{eqnarray}
the Shannon entropy \cite{Sha-48}.
In a sense, $\textbf{p}^{ME}$ is the member of $\mathcal{S}$ having the smoothest distribution.  Objective criteria for its success are of great value, as maximum entropy inference is applied in many contexts.
Using  Lagrange multipliers, it is easy to show that $\textbf{p}^{ME}$ is given formally  by \cite{Jay-57, Pre-13}
\begin{eqnarray}\label{max_ent}
p_i^{ME} = \exp \left [-m -\sum_j \lambda_j f_{ji} \right],
\end{eqnarray}
where $m$ and the $\left \{\lambda_j \right\}$ must now be chosen so that (\ref{max_ent}) satisfies the constraints (\ref{constraints}) and (\ref{normalization}).  As in the $\textbf{p}^{c,1}$ analysis, the normalization condition provides a simple solution for one of the unknowns:
\begin{eqnarray}
e^{m} = \sum_i  e^{- \sum_j \lambda_j f_{ji}}.
\end{eqnarray}
The $\left \{\lambda_j \right \}$ must again be solved for using  (\ref{constraints}).

The distance between $\textbf{p}^{ME}$ and the exact $\textbf{p}^c$ can be estimated analytically by comparing (\ref{max_ent}) to (\ref{centroid_est}), which take a very similar form.   Assuming the $\{f_{ji} \}$  are Gaussian distributed, with $P(f_{ji}) \propto \exp\left [ - f_{ji}^2 / 2 \sigma^2\right ]$,  expanding either  $\textbf{p}^{c,1}$ or $\textbf{p}^{ME}$ to first order in the $\{f_{ji} \}$ results in the following solution:
\begin{eqnarray} \label{leading_order}
p_i \sim \alpha + \sum_j \beta_{j} f_{ji} + \ldots,
\end{eqnarray}
where $\alpha$ and the $\{\beta_j\}$ are given by
\begin{eqnarray} \nonumber \label{cons_values}
\alpha &\sim & \frac{1}{N} \\
\beta_{j} &\sim & \frac{1}{\sum_{k} f_{jk}^2} \times \left \{1 - \frac{\sum_i f_{ji}}{N} \right \},
\end{eqnarray}
the values needed for (\ref{leading_order}) to satisfy (\ref{constraints}) and (\ref{normalization}) to leading order in $N$.  The leading form  (\ref{leading_order}), (\ref{cons_values}) is common to both $\textbf{p}^{c,1}$ and $\textbf{p}^{ME}$ because they both take the form of functions having arguments linear in the $\{f_{ji} \}$.  If $\sigma \gg \sqrt{N}$, the condition formally defining the weak constraint limit for Gaussian-distributed $\{f_{ji} \}$ \footnote{  For random $\textbf{p}$ and for Gaussian distributed $\{f_{ji}\}$,  $\textbf{p} \cdot \textbf{f}_j   \sim O\left ( \frac{\sigma}{N^{1/2}} \right)$.  If this typical dot product value is very large, the constraint  (\ref{constraints}) resembles one of orthogonality, $\textbf{p} \cdot \textbf{f}_j \approx 0$.  On the other hand, if $\frac{\sigma}{N^{1/2}} \ll O(1)$, satisfaction of   (\ref{constraints}) requires $\textbf{p}$ nearly parallel to $\textbf{f}_j$, a much stronger condition. Thus, the ratio $\sigma/N^{1/2}$ determines whether the constraints are strong or weak.},  the term proportional to   $\sum_i f_{ji}$ dominates $\beta_j$ in (\ref{cons_values}), and the second term in (\ref{leading_order}) is of order $O\left (\frac{\mathcal{C}^{1/2}}{N^{3/2}}\right)$.  This is smaller than the leading $\alpha$ contribution in (\ref{leading_order}), which is $O(N^{-1})$.  In this case, the distance between $\textbf{p}^{ME}$ and $\textbf{p}^{c,1}$ can be estimated by considering expansion up to second order in the $\{f_{ji}\}$, where the two solutions have differing Taylor series coefficients:  $e^x \sim 1 + x + \frac{ x^2}{2} + \ldots$, while $\frac{1}{1-x}\sim 1 + x + x^2 + \ldots$.  This gives 
\begin{eqnarray}
p_i^{c,1} - p_i^{ME} \sim O \left ( \sum_j \frac{\beta_{j}^2}{\alpha} f_{ji}^2 \right) \sim O\left (\frac{\mathcal{C}}{N^2} \right),
\end{eqnarray}
much smaller than the width of the solution space, which, from (\ref{width}) and (\ref{leading_order}), is given by $ \sigma_{p_i} \sim O\left (\frac{1}{N}\right)$.  The maximum entropy and centroid solutions are very close  in the large $N$, weak constraint limit. 

In the strong constraint limit, $\sigma \ll N^{1/2}$, the first term in (\ref{cons_values}) dominates $\beta_{j}$, and the second term in (\ref{leading_order}) is $\sum_j \beta_{j} f_{ji} \sim O\left (\frac{1}{N \sigma} \right)$.  As $\sigma \to O(1)$, this is no longer smaller than $\alpha$, signaling the breakdown of the asymptotic expansion.   Empirically, I find that in this case
\begin{eqnarray}
p_i^{c,1} - p_i^{ME} \sim O \left ( \sigma_{p_i} \right ) \sim O \left (p_i^{c,1}\right).
\end{eqnarray}
That is, in the strong constraint limit, the two solutions are distant, with component separations comparable to the solution space widths.  A typical example illustrating this is shown in Fig.\ \ref{fig:snub}.  Here,  as expected,  $\textbf{p}^{c,1}$ is much closer to the exact $\textbf{p}^c$ (obtained via averaging over a random walk through $\mathcal{S}$) than is $\textbf{p}^{ME}$.   The discrepancy between the two is largest when $p_i^{c}$ (which sets the width $\sigma_{p_i}$) is large.   Further, $p^{ME}_i < p^{c}_i$ for all components taking relatively large or relatively small values, whereas $p_i^{ME} > p_i^c$ for all $i$ taking intermediate values.  This qualitative observation appears to hold quite generally, with entropy maximization occurring at a point whose intermediate weight components are substantially bolstered relative to those of the centroid, while all other components are relatively diminished.

In summary, then, I have shown that the centroid $\textbf{p}^c$ of $\mathcal{S}$ provides the formal solution to (\ref{constraints}) and (\ref{normalization}) that minimizes (\ref{av_error}).  Although other variational score functions could be employed -- \textit{e.g.,} the entropy -- (\ref{av_error}) represents a useful one to consider, in that it provides an objective measure for the expected error.  By comparing the popular maximum entropy solution $\textbf{p}^{ME}$ to the centroid's saddle point approximation  -- $\textbf{p}^{c,1}$, given by equations (\ref{saddle_conds})-(\ref{centroid_est}), I have shown that $\textbf{p}^{ME}$ actually performs quite well, in general,  in the weak constraint limit.  This is a very useful result, as most prior tests of the maximum entropy principle have relied upon particular, testable examples.  In the strong constraint limit, the centroid and maximum entropy solutions are distant, and $\textbf{p}^{ME}$ is expected to perform poorly, by measure (\ref{av_error}). In this limit, centroid inference is typically much more accurate. 

 Like maximum entropy inference, centroid inference has the benefit of being free from any bias associated with fitting to a particular, model form.   In practice,  the centroid estimate can be obtained through averaging over a random walk through $\mathcal{S}$.  However, the walk time required increases relatively quickly with $N$.  Alternatively, successive analytic approximations to $\textbf{p}^c$ can be obtained using the method I outline here.  The saddle point approximation $\textbf{p}^{c,1}$  provides a simple, first estimate, very similar in form to $\textbf{p}^{ME}$, that is accurate in the large $N$ limit.   Evaluation of  $\textbf{p}^{c,1}$  provides substantial value, even when not working within the uniform ensemble, as it immediately provides much information relating to the solution set's geometry, as well as to the strength of the applied constraints.

\begin{acknowledgments}
I thank Mike DeWeese for helpful discussions, Frank Brown and Phil Pincus for helpful comments, Jonathan Bergknoff for computer programming assistance, and the USA NSF for support through grant No.\ DMR-1101900.
\end{acknowledgments}

\bibliography{refs}

\begin{thebibliography}{9}
\expandafter\ifx\csname natexlab\endcsname\relax\def\natexlab#1{#1}\fi
\expandafter\ifx\csname bibnamefont\endcsname\relax
  \def\bibnamefont#1{#1}\fi
\expandafter\ifx\csname bibfnamefont\endcsname\relax
  \def\bibfnamefont#1{#1}\fi
\expandafter\ifx\csname citenamefont\endcsname\relax
  \def\citenamefont#1{#1}\fi
\expandafter\ifx\csname url\endcsname\relax
  \def\url#1{\texttt{#1}}\fi
\expandafter\ifx\csname urlprefix\endcsname\relax\def\urlprefix{URL }\fi
\providecommand{\bibinfo}[2]{#2}
\providecommand{\eprint}[2][]{\url{#2}}

\bibitem[{\citenamefont{Shlens et~al.}(2006)\citenamefont{Shlens, Field,
  Gauthier, Grivich, Petrusca, Sher, Litke, and Chichilnisky}}]{Shl-06}
\bibinfo{author}{\bibfnamefont{J.}~\bibnamefont{Shlens}},
  \bibinfo{author}{\bibfnamefont{G.~D.} \bibnamefont{Field}},
  \bibinfo{author}{\bibfnamefont{J.~L.} \bibnamefont{Gauthier}},
  \bibinfo{author}{\bibfnamefont{M.~I.} \bibnamefont{Grivich}},
  \bibinfo{author}{\bibfnamefont{D.}~\bibnamefont{Petrusca}},
  \bibinfo{author}{\bibfnamefont{A.}~\bibnamefont{Sher}},
  \bibinfo{author}{\bibfnamefont{A.~M.} \bibnamefont{Litke}}, \bibnamefont{and}
  \bibinfo{author}{\bibfnamefont{E.~J.} \bibnamefont{Chichilnisky}},
  \bibinfo{journal}{J. Neurosci.} \textbf{\bibinfo{volume}{26}},
  \bibinfo{pages}{8254} (\bibinfo{year}{2006}).

\bibitem[{\citenamefont{Schneidman et~al.}(2006)\citenamefont{Schneidman,
  Berry, Segev, and Bialek}}]{Sch-06}
\bibinfo{author}{\bibfnamefont{E.}~\bibnamefont{Schneidman}},
  \bibinfo{author}{\bibfnamefont{M.~J.~I.} \bibnamefont{Berry}},
  \bibinfo{author}{\bibfnamefont{R.}~\bibnamefont{Segev}}, \bibnamefont{and}
  \bibinfo{author}{\bibfnamefont{W.}~\bibnamefont{Bialek}},
  \bibinfo{journal}{Nature} \textbf{\bibinfo{volume}{440}},
  \bibinfo{pages}{1007} (\bibinfo{year}{2006}).

\bibitem[{\citenamefont{Cocco et~al.}(2009)\citenamefont{Cocco, Leibler, and
  Monasson}}]{coc-09}
\bibinfo{author}{\bibfnamefont{S.}~\bibnamefont{Cocco}},
  \bibinfo{author}{\bibfnamefont{S.}~\bibnamefont{Leibler}}, \bibnamefont{and}
  \bibinfo{author}{\bibfnamefont{R.}~\bibnamefont{Monasson}},
  \bibinfo{journal}{Proc. Natl. Acad. Sc.} \textbf{\bibinfo{volume}{106}},
  \bibinfo{pages}{14058} (\bibinfo{year}{2009}).

\bibitem[{\citenamefont{Press\'{e} et~al.}(2013)\citenamefont{Press\'{e},
  Ghosh, Lee, and Dill}}]{Pre-13}
\bibinfo{author}{\bibfnamefont{S.}~\bibnamefont{Press\'{e}}},
  \bibinfo{author}{\bibfnamefont{K.}~\bibnamefont{Ghosh}},
  \bibinfo{author}{\bibfnamefont{J.}~\bibnamefont{Lee}}, \bibnamefont{and}
  \bibinfo{author}{\bibfnamefont{K.~A.} \bibnamefont{Dill}},
  \bibinfo{journal}{Rev. Mod. Phys.} \textbf{\bibinfo{volume}{85}},
  \bibinfo{pages}{1115} (\bibinfo{year}{2013}).

\bibitem[{\citenamefont{Albanna et~al.}(2012)\citenamefont{Albanna, Hillar,
  Sohl-Dickstein, and DeWeese}}]{alb-12}
\bibinfo{author}{\bibfnamefont{B.~F.} \bibnamefont{Albanna}},
  \bibinfo{author}{\bibfnamefont{C.}~\bibnamefont{Hillar}},
  \bibinfo{author}{\bibfnamefont{J.}~\bibnamefont{Sohl-Dickstein}},
  \bibnamefont{and} \bibinfo{author}{\bibfnamefont{M.~R.}
  \bibnamefont{DeWeese}}, \bibinfo{journal}{arXiv preprint arXiv:1209.3744}
  (\bibinfo{year}{2012}).

\bibitem[{\citenamefont{Rademacher}(2007)}]{rad-07}
\bibinfo{author}{\bibfnamefont{L.~A.} \bibnamefont{Rademacher}}, in
  \emph{\bibinfo{booktitle}{Proceedings of the twenty-third annual symposium on
  Computational geometry}} (\bibinfo{organization}{ACM}, \bibinfo{year}{2007}),
  pp. \bibinfo{pages}{302--305}.

\bibitem[{\citenamefont{Jones}(1943)}]{Jon-43}
\bibinfo{author}{\bibfnamefont{L.~M.} \bibnamefont{Jones}},
  \emph{\bibinfo{title}{An Introduction to Mathematical Methods of Physics}}
  (\bibinfo{publisher}{Benjamin Cummings}, \bibinfo{year}{1943}).

\bibitem[{\citenamefont{Shannon}(1948)}]{Sha-48}
\bibinfo{author}{\bibfnamefont{C.~E.} \bibnamefont{Shannon}},
  \bibinfo{journal}{Bell System Tech. J.} \textbf{\bibinfo{volume}{27}},
  \bibinfo{pages}{379} (\bibinfo{year}{1948}).

\bibitem[{\citenamefont{Jaynes}(1957)}]{Jay-57}
\bibinfo{author}{\bibfnamefont{E.~T.} \bibnamefont{Jaynes}},
  \bibinfo{journal}{Phys. Rev.} \textbf{\bibinfo{volume}{106}},
  \bibinfo{pages}{620} (\bibinfo{year}{1957}).

\end{thebibliography}

\end{document}